 \documentclass[10pt,preprint]{aastex63}
 \usepackage{graphicx,subfigure}

\def\a{\"a}
\def\o{\"o}
\def\u{\"u}
\def\A{\"A}
\def\O{\"O}
\def\U{\"U}

\def\ang{\AA}

\def\gapprox{\lower.4ex\hbox{$\;\buildrel >\over{\scriptstyle\sim}\;$}}
\def\lapprox{\lower.4ex\hbox{$\;\buildrel <\over{\scriptstyle\sim}\;$}}

\shortauthors{Aschwanden}
\shorttitle{Avalanches}

\begin{document}

\renewcommand{\topfraction}{0.95}
\renewcommand{\bottomfraction}{0.95}
\renewcommand{\textfraction}{0.05}
\renewcommand{\floatpagefraction}{0.95}
\renewcommand{\dbltopfraction}{0.95}
\renewcommand{\dblfloatpagefraction}{0.95}

\title{Nonlinear Physics and Self-Organized Criticality 
in Astrophysics}
 
\author{Markus J. Aschwanden}
\affil{Lockheed Martin, Solar and Astrophysics Laboratory (LMSAL),
       Advanced Technology Center (ATC),
       A021S, Bldg.252, 3251 Hanover St.,
       Palo Alto, CA 94304, USA;
       e-mail: markus.josef.aschwanden@gmail.com}

\begin{abstract}
Since standard photon statistics is not applicable 
to nonlinear physics processes, it is of utmost importance
to properly include the nonlinearity in astrophysical
instabilities and emission processes. 
Bak et al.~(1987) postulated sandpile
avalanches and called the
underlying principle ``self-organized criticality'' 
(SOC), but its definition remained  
ambiguous and controversial. Here we discuss the
nonlinear physics of six basic SOC definitions, such as:
(i) phase transitions, (ii) fractal structures, (iii) power laws,
(iv) SOC contingency, (v) exponential growth, and (vi) diffusive
decay. Statistics of SOC events or avalanche processes 
can be inferred from their size distributions, or from 
their spatio-temporal evolution, described with the
logistic equation. Here we quantify each nonlinear 
process in terms of quantitative scaling laws.
\end{abstract}

\section{Introduction}

The concept of {\sl self-organized criticality (SOC)} 
processes has been introduced in statistical mechanics, 
such as phase transitions in terms of the Ising model 
(Christensen and Moloney 2005), and other cellular 
automaton concepts, known as the 
Bak-Tang-Wiesenfeld (BTW) model (Bak et al.~1987, 1988;
Pruessner 2012).
A schematic representation of the range of perception
of SOC is shown in Fig.~1 of Watkins et al.~(2016), which
illustrates four levels of SOC definitions:
The most restrictive definition of a SOC process is 
the self-tuned phase transition, the second level is 
fractality, the third level is power law-like
size distributions, and the fourth level is contingency 
of nature caused by SOC. 
Some of the most common nonlinear physics processes
involve instabilities with an exponential growth phase,
followed by a diffusive decay phase, which may be 
called {\sl ``avalanche events''}.
The schematic shown in Table~1 
indicates that the definition of avalanches or SOC events 
corresponds to at least six different nonlinear physics 
categories.

The parameterization of SOC events is generally
given in terms of a fitted power law slope 
($\alpha_x$) of their size distribution function, 
$N(x) dx \propto x^{-\alpha_x} dx$, 
while the parameterization of an 
avalanche event is generally available from
the time profile $x(t)$ of a single event.
For instance, the SOC parameter of energy $x=E$
characterizes the power law slope
$\alpha_E$, while the time evolution $E(t)$ of
each event reveals information on the exponential 
growth phase of an instability.

In this study we describe six processes that are
relevant for the definition of nonlinear physics
processes, which provide statistics of SOC events
or avalanche events. In principle, each of the 
six nonlinear physics processes can be used as
an individual SOC definition or model, while the 
combination of all six models attempts a more 
comprehensive approach. 

\begin{table}[h]
\begin{center}
\caption{Nomenclature of nonlinear physics processes
(adapted from Watkins et al.~2016).}
\medskip   
\begin{tabular}{|l|l|l|}
\hline
PARAMETERS :              & NONLINEAR PHYSICS :   & EVENT STATISTICS : \\
\hline
\hline
size distributions $\alpha_x$ & phase transition  & Self-organized criticality	\\
	                  & fractality            & (SOC) events \\
	                  & power laws            &	\\
	                  & contingency	          &	\\
\hline
spatio-temporal evolution $x(t)$ & exponential growth & avalanche events	\\
	                  & diffusive decay 	  &	\\
\hline
\end{tabular}
\end{center}
\end{table}

\section{Discussion}

In the following we discuss the six nonlinear physics 
categories enumerated in Table 1. For each of the six
SOC processes we provide a brief description of the
underlying physics and a quantative parameterization 
that can be used as a SOC definition for observed or 
predicted astrophysical phenomena.

\subsection{Self-tuned phase transitions}

{\sl Criticality refers to the behavior of exended
systems as a phase transition where no chacteristics
scale exists. Thermodynamically, a phase transition
occurs when there is a singularity in the free energy.
The liquid-gas, conductor-superconductor, 
fluid-superfluid, or paramagnetic-ferromagnetic
phase transition are common examples.
The Ising model of a ferromagnet
is one of the simplest models displaying the
paramagnetic-ferromagnetic phase transition, that is,
the spontaneous emergence of magnetization in zero
external field as the temperature is lowered below
a certain critical temperature. At the critical
point of critical temperature and zero external field
there is no characteristic scale}
(Christensen and Moloney 2005). 

A definition of the Ising model can be given 
by a two-dimensional grid (in two dimensions),
$N=L \times L$,
where every site is occupied by a so-called spin $s_i$.
For a magnetic material, one may think of the spins
as the magnetic dipoles positioned on the crystal
structure lattice. In uni-axial magnetic materials,
the magnetic dipole interactions constrain the spins to
point in parallel ($s_i=+1$) or anti-parallel direction
($s_i=-1$). In the Ising model, the total internal 
interaction energy $E_{int}$ is quantified by the coupling
constants between the spins $s_i$ and $s_j$, and
the sum runs over all distant pairs of spins.
In addition, the total energy $E_{ext}$ is characterized
by an external field, where the sum $H$ runs only over the
nearest neighbors, $s_{i}$ and $s_{j}$, so that
the combined internal and external contributions add up to
(Ising 1925),
\begin{equation}
	E_{s_i} = E_{int} + E_{ext} 
		= - J \sum_{<ij>} s_i s_j - H \sum_{i=1}^N s_i \ .
\end{equation}
where $H$ is the uniform external field and $<i,j>$ is
a shorthand notation for the full microscopic desription
of a particular microstate.

The behavior of the Ising model as a function of the
temperature, defined by the energy scale $k_B T$ 
(with the Boltzmann's constant $k_B$), the spin-spin 
interaction $k_B T \ll J$ is relatively strong,
so that the spin tends to align with one another,
reducing the system's internal energy. For temperatures
$k_B T \gg J$, the spin-spin interaction is relatively
weak, so that the spins are effectively non-interacting
and point up and down randomly. The unpredictable 
aggregation of small and large magnetic (spin) domains 
in the Ising model became a paradigm of self-organized 
criticality. The Ising model was also applied to biology, 
neuroscience, economics, and sociology (see references 
in Christensen and Moloney 2005, chapter 2).

A theoretical predicition that could be tested is the
fractal dimension $D_A$ in ferro-magnetic domains.
If the fractal dimension is measured in 2-dimensional
areas, the fractal (or Euclidean) dimension would be 
$D_A=2.0$, while curvi-linear phenomena should show a 
fractal (or Euclidean) dimension
of $D_A=1.0$, or a mean value of the fractal dimension
$D_A=(D_{A,min}+D_{A,max})/2=(1+2)/2=1.5$.
As an example, a critical behavior of the Ising model 
on fractal structures has been found in dimensions 
between one and two (Monceau and Perreau, 2001).

\subsection{Fractality}

The seminal SOC paper (Bak et al.~1988) 
states the following role of
fractal structures in SOC processes: {\sl We show 
that certain extended dissipative dynamical systems 
naturally evolve into a critical state, with no 
characteristic time or length scales. The temporal 
``fingerprint'' of the self-organized critical state 
is the presence of flicker noise or 1/f noise; its 
spatial signature is the emergence of scale-invariant 
(fractal) structure} (Bak et al.~1987, 1988).

While the Euclidean relationship $A \propto L^2$
implies non-fractal areas, the fractal area $A$ 
can be expressed in terms of a power law index $D_A$ 
of the length scale $L$, also called the Hausdorff
dimension (Mandelbrot 1977, 1983, 1985),
\begin{equation}
	A = L^{D_A} \ .
\end{equation}
Likewise, while the Euclidean relationship $V \propto L^3$
implies non-fractal volumes, the fractal volume $V$ 
can be expressed in terms of a power law index $D_V$ 
of the length scale $L$, also called the Hausdorff
dimension, 
\begin{equation}
	V = L^{D_V} \ .
\end{equation}
There is no universal theory that predicts the values
of the fractal dimensions, but some educated guess
can be made by assuming the Euclidean dimension 
$d_{max}=(d)$ as upper limit, and one dimension less, 
$d_{min}=(d-1)$, as lower limit. This yields the 
predictions,
\begin{equation}
	D_A ={(D_{A,min}+D_{A,max}) \over 2} 
	    ={1+2 \over 2} = {3 \over 2} = 1.5 \ 
\end{equation}
and
\begin{equation}
	D_V ={(D_{V,min}+D_{V,max}) \over 2} 
	    ={2+3 \over 2} = {5 \over 2} = 2.5 \ .
\end{equation}
Since the values $D_A$ and $D_V$ are observables, their
statistical means and standard deviations can be compared
with the theoretical predictions (Eq.~4, 5). Low values 
of the fractal dimension $D_A$ are characteristic for 
curvi-linear geometries (d=1), while high values of the
fractal dimension $D_V$ indicate Euclidean geometries
with solid filling (d=3). This dimensional diagnostic 
yields information on the spatial topology of
astrophysical structures. Magnetic field lines tend
to display 1-dimensional geometries, while 3-dimensional
plasma structures reveal solidly-filled 3-dimensional 
volumes. 

2-dimensional structures have been observed
and theoretically modeled as quasi-linear separatrix
surface layers in magnetic reconnection processes 
(e.g., Aulanier et al.~2005).
Multi-fractal solar EUV intensity fluctuations were found
to be able to discriminate between coronal heating models
(e.g., Cadavid et al.~2016).
An example of fractal structures is shown in Fig.~1 that
yields a mean value and standard deviation of 
$D_A=1.55\pm 0.03$, observed
in a large solar flare. This is close to the theoretical
prediction of $D_A=1.5$ (Aschwanden and Aschwanden 2008a).
Similar values were obtained from sunspots ($D_A=1.54\pm0.13$)
and from active regions ($D_A=1.59\pm0.20$),
while photospheric and transition region phenomena
reveal distinctly lower values ($D_A=1.21\pm0.07$ and
$D_A=1.26\pm0.13$) (Aschwanden and Nhalil 2022).

\subsection{Power laws}

Many avalanche processes have been interpreted in 
terms of {\sl ``self-organized criticality (SOC)} 
processes, which often display power law size
distributions (Charbonneau et al.~2001;
Aschwanden 2011, 2012, 2014, 2022; Aschwanden et al.~2016; 
McAteer et al.~2016; Sharma et al.~2016).
Self-organized criticality phenomena ubiquitously display
power laws in their size distribution functions,
which can be analytically derived for each SOC parameter.
We need only two fundmental assumptions: (i) the universal
scale-free probablity distribution function 
$N(L) dL$ of length scales $L$,
which simply expresses a reciprocal relationship between the
number density distribution $N(L)$ and the fractal volume $L^{d}$, 
\begin{equation}
	N(L)\ dL = L^{-d} \ dL \ ,
\end{equation}
and (ii) a scaling law relationship beetween 
the length scale $L$ and a SOC parameter, for instance
the volume $(V)$, 
\begin{equation}
	V(L) = L^{(D_V)} ,
\end{equation}
which can be obtained from the
fractal dimension $D_V$, where $d=1,2,3$ are the
Euclidean dimensions. A dimension $d=3$ is often used
for three-dimensional data, but other dimensions $d=2$ 
can also be useful for fractal areas. In order to
derive the power law of (fractal) volumes we invert
the SOC parameter, $L \propto V^{(1/D_V)}$, and
calculate its derivative, $dL/dV=V^{(1/D_V)-1}$,
which yields,
\begin{equation}
	N(V) dV = L(V)^{-d}\ | {dL / dV} | dV 
	= V^{-\alpha_V} dV \ .
\end{equation}
and the desired power law index $\alpha_V$ is,
\begin{equation}
	\alpha_V = 1 + {(d-1) \over D_V} \ ,
\end{equation}
which yields for $d=3$ and $D_V=5/2$ a value of
$\alpha_V=9/5 = 1.80$.
Similar derivations of power law indices can be
obtained for other SOC parameters, such as for
areas $A$, fluxes $F$, energies $E$, or fluences 
(see Aschwanden 2025). An exhaustive list of power
law slope measurements in energies $\alpha_E$ and fluxes
$\alpha_F$ are compiled in a recent textbook,
covering astrophysical phenomena such as solar and
stellar flares, magnetospheric events, planetary
systems, galactic and black-hole systems
(Aschwanden 2025).

\subsection{SOC Contingency}

Watkins et al.~(2016) considered the contingency of
nature caused by SOC as the most visionary perception
of SOC phenomena. Generally,
the aspect of contingency is used in the context of
an event (i) that may occur but that is not likely
or intended, (ii) a possibility that must be prepared for,
or (iii) the condition of being dependent on chance
(Wikipedia).

In the paper "Complexity, Contingency, and Criticality"
by Bak and Paczuski (1995) it is stated that:
{\sl Complexity originates from the tendency of large 
dynamical systems to organize themselves into a 
critical state, with avalanches or "punctuations" 
of all sizes. In the critical state, events which 
would otherwise be uncoupled become correlated. 
The apparent, historical contingency in many sciences, 
including geology, biology, and economics, finds a 
natural interpretation as a self-organized critical 
phenomenon. These ideas are discussed in the context 
of simple mathematical models of sandpiles and 
biological evolution. Insights are gained not only 
from numerical simulations but also from rigorous 
mathematical analysis.}

In practice, contingency may be used as a strategy
to distinguish nonlinear physics from linear physics.
Linear physics is traditionally applied to random
processes, while nonlinear physics invokes a ``fat tail''
in the size distribution of SOC events. The statistics
of event sizes in linear physics applications can be
approximated with a Poissonian, an exponential, or a Gaussian
size distribution, while nonlinear physics requires
generally a power law-like distribution. The practical 
goal of any contingency measure is the estimation of the 
largest possible event.  If the size distribution is
characterized with a SOC power law slope $\alpha_V$, 
i.e., $N(V) dV \propto V^{-\alpha_V} dV $, the largest
event (or contingency measure), is
\begin{equation}
	N(V=V_{max}) dV = (V_{max})^{-\alpha_V} dV \ ,
\end{equation}
where $V_{max}$ is the maximum volume of a SOC event,
while $L_{max} \propto V_{max}^{(1/D_V)}$ corresponds 
to the maximum length scale. The maximum  length
scale $L_{max}$ is also called the {\sl finite system
size}, which is detectable by a steep cutoff in the
size distribution. Of course, the maximum event of a
SOC size distribution also depends on the number of events
in the size distribution. 

In order to give some examples, the maximum size 
$L_{max}$ corresponds to the size of continental 
tectonic plates in the case of earthquakes,
i.e., $L_{max} \approx R_E = 6370$ km.
In the case of solar (or stellar) flares, the 
size of (magnetic) active regions size amounts to
$L_{max}\approx 200 \times 10^8$ m, 
Aschwanden and Aschwanden 2008a, 2008b).
These limits can be considered as reliable 
contingency measures.
 
\subsection{Exponenial Growth - Logistic Avalanches}

An instability in astrophysics can be formulated with
an exponential growth phase and a subsequent diffusion 
phase. The initial growth phase can be expressed with 
a first-order differential equation,
\begin{equation}
	{dE(t) \over dt} \approx {E(t) \over T} \ ,
\end{equation}
where $t$ is the time parameter, and
$T$ is the exponential growth time scale, 
also known as {\sl e-folding time scale},
and $E(t)$ is the time evolution of an energy parameter.

In order to avoid exponential growth to infinity,
which is an unphysical property,
it is appropriate to introduce a factor of
$1-E(t)/E_\infty$, which leads to the so-called
logistic equation, 
\begin{equation}
	{dE(t) \over dt} \approx {E(t) \over T} 
	\left( 1 - {E(t) \over E_\infty} \right) \ ,
\end{equation}
discovered by Piere Francois Verhulst in 1845 
(Beltrami 1987, pp.61-64; Aschwanden et al.~1998).
The logistic equation represents a generalization of the
exponential growth curve (Eq.~11), by including
the growth limitation at the asymptotic saturation 
level $E_\infty$. The solution $E(t)$ of this
differential equation is initially exponental
(because the term $[1-E(t)/E_\infty]$ is close
to unity for $E(t) \ll E_\infty$), and approaches 
asymptotically the value $E_\infty$ for large
times. The physical basis of the logistic term
$[1-E(t)/E_\infty]$ is the constraint of an
uppermost limit $E_\infty$, which is also called
{\sl carrying capacity} in ecology sciences
(e.g., Beltrami 1987; Aschwanden
et al.~1998).

The exact solution of the first-order differential
equation (Eq.12) is,
\begin{equation}
	E(t) = {E_\infty \over 1 + \exp[-(t-t_0)/T]} \ ,
\end{equation}
where $t_0$ is the reference time.
We see that the time evolution of the energy $E(t)$
exhibits a monotonic function in the range of
$0 \le E(t) \le E_\infty$ (Fig.~2a).

The most directly observed parameter of an
astrophysical source is given in flux units $F(t)$, 
(in terms of physical cgs units [erg s$^{-1}$]),
while the energy $E(t)$ is defined by the time integral
of the flux, $E(t)=\int dF(t) dt$, or vice versa
by the time derivative $F(t)=dE(t)/dt$. 
The flux can be expressed in explicit form $F(t)$,
\begin{equation}
	F(t) = {dE(t) \over dt}
	     = {E_\infty \over T} \
	             {\exp[-(t-t_0)/T] \over 
	       \{ 1 + \exp[-(t-t_0)/T]\}^2} \ .
\end{equation}
This explicit form of the energy as a function of time
is shown in Fig.(2b), which ressembles to a Gaussian
function. Herewith we have explicit functions of the
enegy $E(t)$ (Eq.~13) and the flux $F(t)$ (Eq.~14), 
parameterized by three variables, i.e., the finite 
energy maximum $E_\infty$ and the time scale $T$, 
as well as the reference time $t_0$. 
The exponential growth dynamics
predicts spatio-temporal functions for the
energy $E(t)$ and the flux $F(t)$, which can be
tested for each avalanche event separately.

Four examples of logistic curve fits are
shown in Fig.~(3), after subtraction of a linear
background. The solar flare of 1991-May-31
featured in Fig.~3 displays a hierarchy of
``elementary'' time profiles of the hard X-ray flux, 
which fit the logistic equation well, and thus
underscore the nonlinear avalanche behavior
during the rise time, and subsequent diffusion
during the decay phase.

The property of exponential growth is also
called {\sl coherent growth}, in contrast
to the case of {\sl incoherent growth},
which does not exceed the random noise level.
The predicted exponential growth phase 
represents the most obvious manifestation
of self-organized criticality, especially
because each SOC event amplifies nonlinear
growth by several order of magnitudes,
as expected for models of sandpile avalanches.

\subsection{Diffusive Decay}

As the previous description of avalanches governed
by exponential growth implies, accumulation of
flux or energy leads to a monotonic increase of
energy during the rise phase and subsequent
diffusion during the subsequent decay phase.  
It was found that the coefficient of turbulent
diffusion changes in direct proportion to both
the temporal and spatial scales 
(e.g., Abramenko et al.~2011).
Statistics of SOC events can be separately
obtained during the rise time 
$t_{rise} \approx [t_0-T/2, t_0]$
and the decay time 
$t_{decay} \approx [t_0, t_0+T/2]$, since
the two evolutions are likely to be dominated
by two different nonlinear physics processes,
i.e., exponential growth during the rise time,
and diffusion during the decay time.
Consequently, the time-integrated fluxes $F(t)$
in the two time domains are expected to be
correlated,
\begin{equation}
	E_{rise} \approx \int_{t_0-T/2}^{t_0}\ F(t) dt \ ,
\end{equation}
and
\begin{equation}
	E_{decay} \approx \int_{t_0}^{t_0+T/2}\ F(t) dt \ ,
\end{equation}
which yields the prediction,
\begin{equation}
	E_{rise} \propto E_{decay} \ .
\end{equation}
If a proportionality between these two energy parameters
is observed, the (nonlinear) physical processes are
predicted to be incoherent, such as X-ray bremsstrahlung 
and gyrosynchrotron emission in some astrophysical SOC
processes.

\section{Conclusions}

There is no unique definition of self-organized criticality
(SOC), but we identify at least six nonlinear physics 
processes that may serve as a SOC definition and provide
theoretical predictions. The various nonlinear physics
categories complement (and overlap) with each other 
(see also Fig.~3 in Watkins et al.~2016). We conclude
that the various nonlinear physics categories can be used
to test the agreement between the observed SOC phenomena
and theoretical SOC predictions, following the schematic
described in Table 1 and in the Sections 2.1 to 2.6:

\begin{enumerate}
\item{{\bf Self-tuned phase transitions:} The Ising model
predicts an initially randomly distributed grid of
positive and negative spins to produce ordered magnetic domains
with equal spin directions in a ferromagnet, after the
temperature has self-tuned to a particular critical
value. A theoretical predicition that could be tested
is the fractal dimension $D_A=1.5$ in ferromagnetic
domains, as defined in Eq.~(4) for an Euclidean dimension
$d=2$.}

\item{{\bf Fractality:} The Hausdorff fractal dimension 
in 2-dimensional space, $A = L^{D_A}$, and  
in 3-dimensional space, $V = L^{D_V}$, respectively,
are predicted to have mean values of $D_A=1.5$ and $D_V=2.5$,
which can be tested from imaging data. 
For area parameters $A$, each image pixel has a value
$F_{ij}=1$ or 0, while 3-D fluxes are obtained
by including the line-of-sight component, $F_{ijk}$.
Deviations from the theoretically predicted fractal
dimensions may reveal more complex topologies than the
mean values of $D_A=1.5$ and $D_V=2.5$.}

\item{{\bf Power laws:}}
SOC processes predict not only the mathematical form
of size distribution functions $N(x) dx$ for each
SOC parameter, $x=[L,A,V,F,E]$, but also their
specific power law slope value $\alpha_x$. 
The physical basis of size
distribution functions is the universal scale-free
probability distribution (Eq.~6) and the choice of the fractal 
dimension $D_d$, which depends on the dimensionality
$d$ of the Euclidean geometry. For a dimensionality
of $d=2$, the mean fractal dimension is $D_d=D_2=D_A=d-1/2=3/2$,
and for a dimensionality 
of $d=3$, the mean fractal dimension is $D_d=D_3=D_V=d-1/2=5/2$.
If we adopt the mean fractal dimension as a representative
value for $D_d=(d-1/2)$, the power law slope $\alpha_d$ depends 
on the dimensionality $d$ as
\begin{equation}
	\alpha_d = 1 + {(d-1)\over (d-1/2)} \ .
\end{equation}

\item{{\bf SOC contingeny:} A contingency measure of SOC
processes is the maximum value of length scales $L_{max}$ 
that are observed in a size distribution $N(L)$, which is
also called the finite system size limit. The corresponding
SOC parameter for the maximum volume follows from (Eq.~7), 
i.e., $V_{max}=[L_{max}]^{(D_V)}$.}
	
\item{{\bf Exponential growth - Logistic equation:}
The exponential growth phase of the spatio-temporal
evolution of a SOC process predicts the time evolution
of the energy $E(t)$ (in normalized units $T=1$, and $t_0=0$),
\begin{equation}
	E(t) = { E_\infty \over 1 + e^{-t}} \ , 
\end{equation}
and the flux $F(t)$,
\begin{equation}
        F(t) = E_\infty {e^{-t} \over (1 + e^{-t})^2} \  ,
\end{equation}
which can be compared with the observed flux $F(t)$ and
energy values $E(t)$ in Fig.~(3). 
The histogrammed time profiles shown in Fig.~3
(in subpanels A, B, C, and D) represent 
the observed data $F_i=F(t_i)$, while the
smoothed background-subtracted curves render the
best fits of the logistic equation $F(t)$ (Eq.~20). 
The exponential growth behavior of the logistic equation fit
confirms the role of nonlinear instabilities most directly.}

\item{{\bf Diffusive decay:} 
The energy $E_{rise}$ during the growth phase and the
energy $E_{decay}$ during the diffusive decay phase
may belong to two different physical processes.
A linear regression analysis may reveal whether
the two physical processes are correlated or not.
In solar (or stellar) flares, the rise time is
dominated by nonthermal hard X-ray emission,
while the diffusive decay time is mostly produced 
by the conductive loss rate and the radiative loss rate.
Fitting the exponential growth phase (with Eq.~20)
yields the finite energy $E_\infty$ of a SOC avalanche
and its time duration $T$, but it is still an open
question whether the diffusive decay phase is an
dfferent SOC process or not. Nevertheless, 
the logistic equation satisfactorily fits
the observed data during both the rise time and the
decay time, which confirms the role of a nonlinear 
diffusion process for both the exponential growth
and the diffusive decay phases.}
\end{enumerate}

These six conclusions demonstrate that information on
at least six nonlinear physics processes is available
that has not much been used in the context of astrophysical
phenomena. At the same time, each of the six nonlinear
physics categories can be considered as a possible
definition of a SOC process. However, we should not restrict
to a single definition, unless each of the six models
can be explained with a single universal model.
A complete SOC model should also be able to explain
the differences in power law slopes of various
astrophysical phenomena. Relevant efforts are in the works.

\section*{References}
\def\ref#1{\par\noindent\hangindent1cm {#1}}

\ref{Abramenko, V.I., Carbone, V. and Yurchyshyn, V. 2011,
	ApJ 743/2, article id. 133, 9pp,
	{\sl Turbulent diffusion in the photosphere as
	derived in the photospheric bright point motion}}
\ref{Aschwanden, M.J., Dennis, B.R., and Benz, A.O.
 	1998, ApJ 497, 972-993,
 	{\sl Logistic avalanche processes, elementary 
	time structures, and frequency distributions 
	in solar flares}}
\ref{Aschwanden, M.J. and Aschwanden, P.D.~2008a,
	{\sl Solar flare geometries: 1. The area fractal
	dimension}, ApJ 674, 530-543}
\ref{Aschwanden, M.J. and Aschwanden, P.D.~2008b,
	{\sl Solar flare geometries: 2. The volume fractal
	dimension}, ApJ 674, 544-553}
\ref{Aschwanden, M.J.
        2011, Self-organized criticality in astrophysics. The
        statistics of nonlinear processes in the universe,
	Springer/PRAXIS: Berlin}  
\ref{Aschwanden, M.J. 2012, AA 539, A2,
	{\sl A statistical fractal-diffusive avalanche model
	of a slowly-driven self-organized criticality system}}
\ref{Aschwanden, M.J. 
        2014, ApJ 782, 54, 
        {\sl A macroscopic description of self-organized 
	systems and astrophysical applications}}
\ref{Aschwanden, M.J.
 	2022, ApJ 934:33 (27pp), 
 	{\sl The fractality and size distributions of 
	astrophysical self-organized criticality systems}}
\ref{Aschwanden, M.J.
        2025, Power Laws in Astrophysics. 
	Self-Organzed Criticality Systems, 
	Cambridge University Press: Cambridge.}
\ref{Aschwanden, M.J. and Nhalil, N.V. 2022,
	{\sl Interface region imaging spectrograph (IRIS)
	observations of the fractal dimension in the 
	solar atmosphere}, Frontiers in Astronomy 
	and Space Sciences 9, id. 999319}	
\ref{Aulanier,G., Pariat,E., and Demoulin,P. 2005,
 	{\sl Current sheet formation in quasi-separatrix 
	layers and hyperbolic flux tubes}, AA 444, 961-976}
\ref{Bak, P., Tang, C., and Wiesenfeld, K.
 	1987, PhRvL 59/4, 381-384,
 	{\sl Self-organized criticality: 
	An explanation of 1/f noise}}
\ref{Bak, P., Tang, C., and Wiesenfeld, K.
 	1988, PhRvL 38/1, pp.364-374.
 	{\sl Self-organized criticality}}
\ref{Bak, P. and Paczuski, M. 1995,
	{\sl Complexity, contingency, and criticality},
	Proc. National Academy of Sciences of the
	United States of America 92, 6689-6696.}
\ref{Beltrami, E. 1987,
 	{\sl Mathematics for dynamic modeling},
 	Academic Press, Inc, Boston.}
\ref{Cadavid, A.C., Rivera, Y.J., Lawrence, J.K., et al. 2016,
	ApJ 831/2, article id. 186, 18pp,
	{\sl Multifractal solar EUV intensity fluctuations
	and their implications for coronal heating models}}
\ref{Charbonneau,P., McIntosh,S.W., Liu,H.L., Bogdan,T.J. 2001,
 	Solar Phys. 203, 321-353,
 	{\sl Avalanche models for solar flares}}
\ref{Christensen, K. and Moloney, N.R. 2005,
	Complexity and Criticality, Imperial College Press, 
	London, UK.}
\ref{Ising E. 1925, Beitrag zur Theorie des Ferromagnetismus.
	Z. Phys. 31:253.}
\ref{Mandelbrot, B.B. 1977,
        {\sl Fractals: form, chance, and dimension, 
	Translation of Les objects fractals, 
	W.H. Freeman, San Francisco.}}
\ref{Mandelbrot, B.B. 1983, {\sl The fractal geometry of
	nature}, W.H. Freeman, San Francisco.}
\ref{Mandelbrot, B.B. 1985, {\sl Self-affine fractals
	and fractal dimension}, {\sl Physica Scripta 32,
	257-260}}
\ref{McAteer,R.T.J., Aschwanden, M.J., Dimitropoulou, M., 
	Georgoulis, M.K., Pruessner, G., Morales, L., Ireland, J., 
	and Abramenko, V. 2016, SSRv 198 217-266,
 	{\sl 25 Years of Self-Organized Criticality: 
	Numerical Detection Methods}}
\ref{Monceau, P. and Perreau, M. 2001,
	Phys.Rev.B63, 184420,
	{\sl Critical behavior of the Ising model on fractal 
	structures in dimensions between one and two: 
	Finite-size scaling effects}}
\ref{Pruessner, G. 2012,
 	{\sl Self-organised criticality. Theory, models and 
	characterisation}}
\ref{Sharma,A.S., Aschwanden, M.J., Crosby, N.B., Klimas, A.J., 
	Milovanov, A.~V., Morales, L., Sanchez, R., and Uritsky, V. 2016,
 	SSRv 198, 167-216,
 	{\sl 25 Years of Self-Organized Criticality: 
	Space and Laboratory Plamsas}}
\ref{Watkins, N.W., Pruessner, G., Chapman, S.C., 
	Crosby, N.B., and Jensen, H.J. 2016,
 	SSRv 198, 3-44, 
 	{\sl 25 Years of Self-Organized Criticality: 
	Concepts and Controversies}}

\clearpage

\begin{figure} 
\centerline{\includegraphics[width=1.0\textwidth]{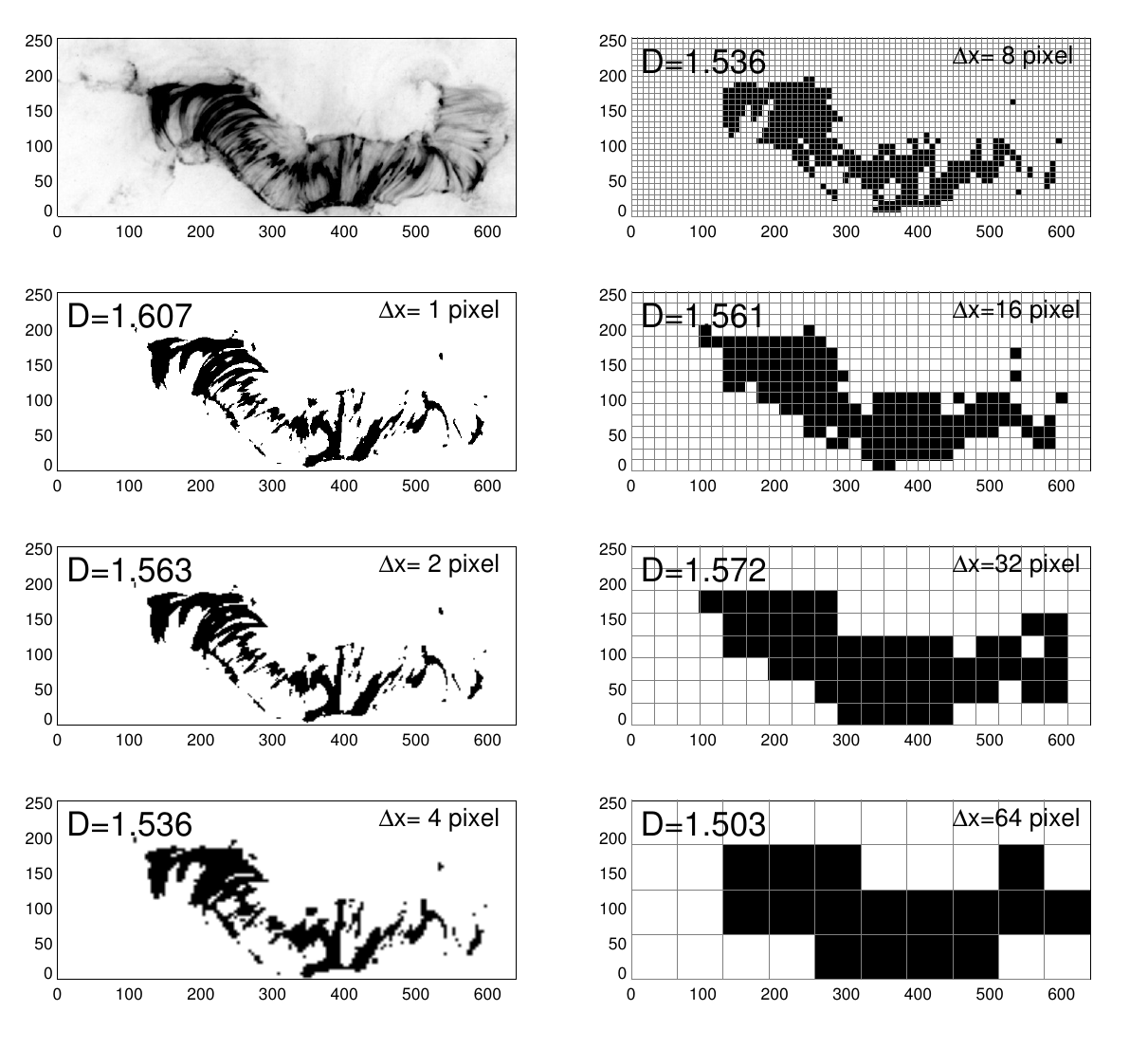}}
\caption{Measurement of the fractal area of a solar flare,
observed by the spacecraft TRACE 171\ang\ on 2000 July 14,
10:59.32 UT. The Hausdorff dimension is evaluated with a
box counting algorithm for pixels above a threshold of
20\% of the peak flux value, yielding a mean of $D_A
=1.55\pm0.03$ for the 8 different spatial scales
(Aschwandenk and Aschwanden 2008a).}
\end{figure}

\begin{figure} 
\centerline{\includegraphics[width=1.0\textwidth]{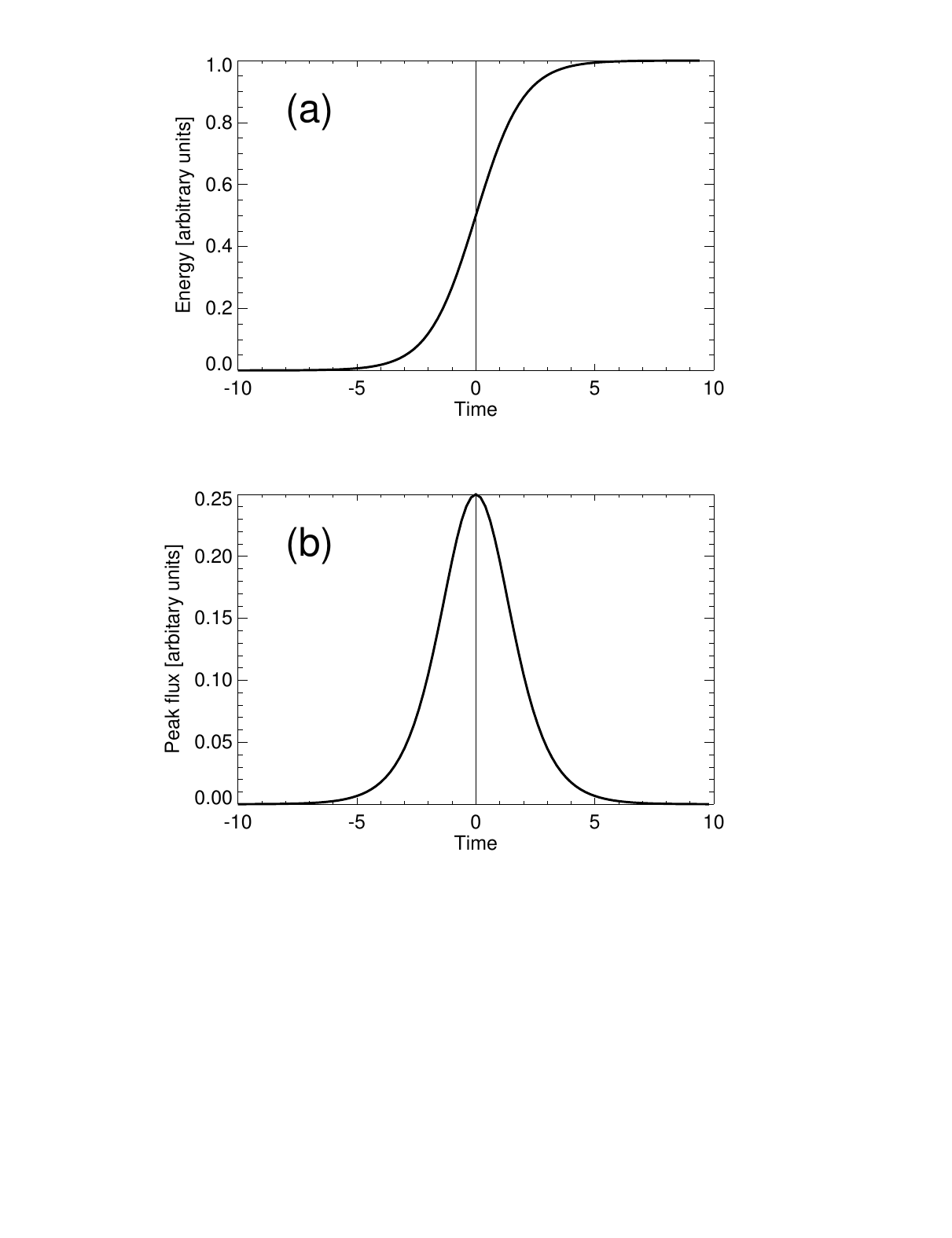}}
\caption{Exact solution of the energy function $E(t)$ (a), and the (peak) 
flux function $F(t)$ of an avalanche event (b).}
\end{figure}

\begin{figure} 
\centerline{\includegraphics[width=1.0\textwidth]{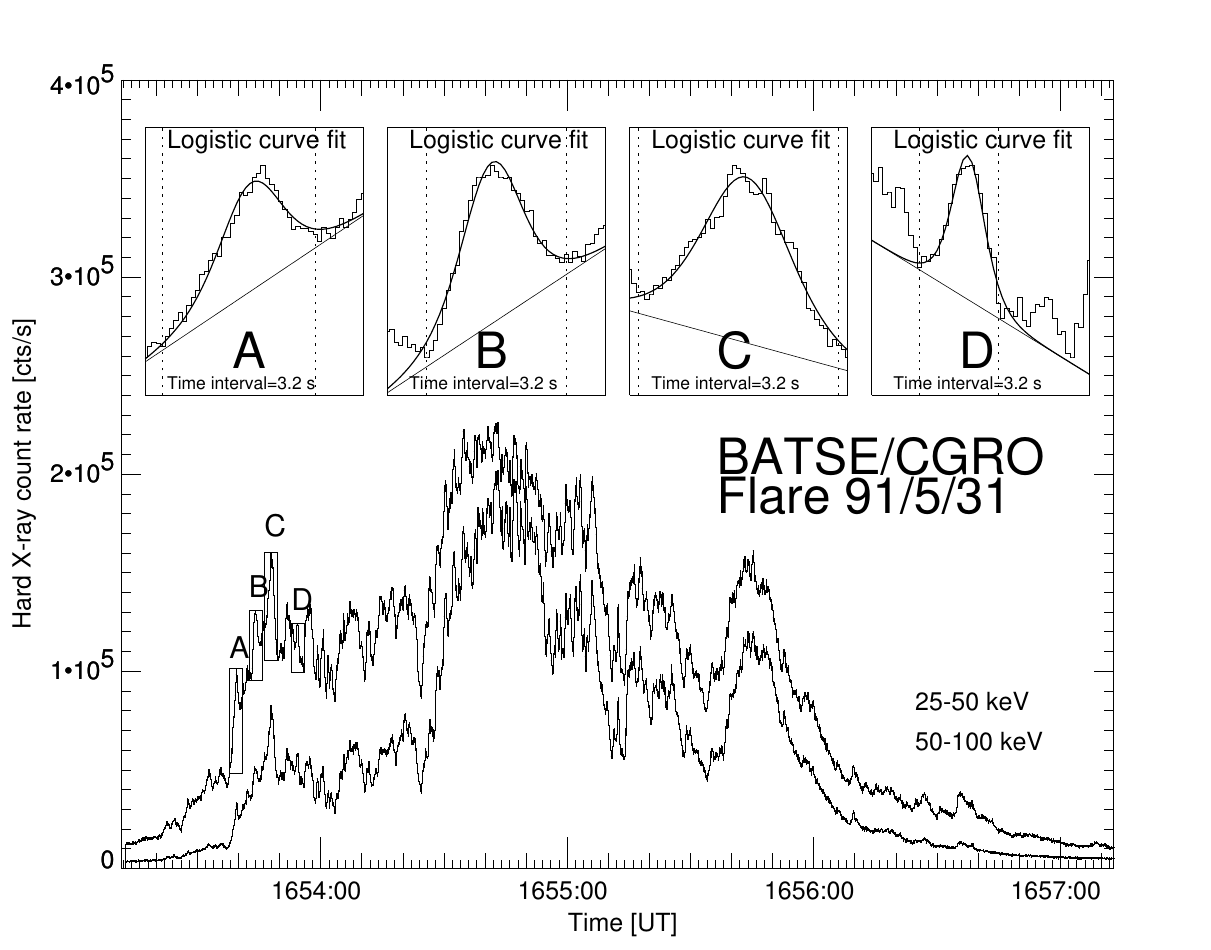}}
\caption{Examples of $>25$ and $>50$ keV hard X-ray time profiles, observed
with BATSE/CGRO 
with 64-ms time resolution. Four sub-bursts at the beginning
of the flare are enlarged (marked with the labels A, B, C, and D.) Note that
the solar flare (1991, May 31) consists of significant sub-bursts with
sub-second time scales, while the sub-bursts themselves show no
significant sub-structures, and thus may represent time-resolved
elementary structures of particla acceleration processes. Four of the
sub-bursts are fitted with the logistic equation, after 
subtraction of a linear background each (Aschwanden et al.~1998).}
\end{figure}

\end{document}